\documentclass[12pt]{iopart}

\usepackage{bm}
\usepackage[dvips]{graphicx}
\usepackage[all]{xy}

\usepackage[dvipdfmx, bookmarks, colorlinks=true, breaklinks]{hyperref}  
\hypersetup{linkcolor=blue,citecolor=blue,filecolor=black,urlcolor=blue} 

\begin{document}

\title{Band structure of hydrogenated Si nanosheets and nanotubes.}
\author{G. G. Guzm\'{a}n-Verri$^1$ and L. C. Lew Yan Voon$^2$}
\address{$^1$ Department of Physics and Astronomy, University of
 California at Riverside, Riverside, California 92521, USA}
\address{$^2$ Department of Physics, Wright State University, Dayton, Ohio 45435, USA }
\ead{gguzm004@ucr.edu and lok.lewyanvoon@wright.edu}

\begin{abstract}
The band structure of fully hydrogenated Si nanosheets and nanotubes
are elucidated by the use of an empirical tight-binding model.
The hydrogenated Si sheet is a semiconductor with  
indirect band gap of about $2.2$\,eV. 
The symmetries of the wave functions allow us to explain the
origin of the gap.
We predict that, for certain chiralities, hydrogenated Si nanotubes represent
a new type of semiconductor, one with co-existing direct and indirect gaps of exactly the same magnitude.
This behavior is different from the Hamada rule established for 
non-hydrogenated carbon and silicon nanotubes. Comparison to an ab initio calculation is made.
\end{abstract}

\maketitle

\section{Introduction}

Graphene exhibits electronic properties that conventional metals do not show, e.g., 
charge carriers with zero effective mass, anomalous quantum Hall effect, 
and a minimum conductivity when carrier concentrations tend to zero~\cite{Geim05a}. 
The unconventional electronic properties of graphene occur due to its linear energy 
dispersion in the vicinity of the Dirac point and two-dimensional structure~\cite{Geim05a, CastroNeto09a}.
Recent theoretical studies show, however, that a linear energy dispersion relation 
is not unique to graphene~\cite{Guzman07a,Cahangirov09a,Sahin09a,Lebegue09a}. 
For instance, a flat two-dimensional silicon (Si) sheet with a honeycomb lattice~(silicene) 
exhibits a linear energy dispersion too~\cite{Guzman07a,Cahangirov09a,Sahin09a, Lebegue09a}.
Even buckling of the Si honeycomb lattice~(Si tends to form $sp^3$ bonds rather than $sp^2$ bonds)
does not remove the linearity in the energy dispersion relation nor does it open 
an energy gap~\cite{Guzman07a}. More generally, it has been predicted that a 
group-IV nanosheet with a flat or buckled honeycomb lattice also shows a linear dispersion 
too~\cite{Cahangirov09a, Sahin09a, Lebegue09a, Lok10a}. 
It is the underlying symmetries of the honeycomb lattice that are responsible 
for the linear energy dispersion~\cite{Guzman07a}.
One-dimensional carbon nanostructures also exhibit unconventional electronic properties, e.g., 
carbon nanotubes show different electronic behavior depending on their chirality. 
Similarly to graphene, the unconventional electronic properties are not unique 
to carbon nanotubes~\cite{Guzman07a, Ciraci05a}.
Si nanotubes also show electronic properties that are chirality dependent~\cite{Guzman07a, Ciraci05a}.

The theoretical studies mentioned above motivated the experimental realization of 
two-dimensional sheets other than graphene. 
Silicon hexagonal sheets were chemically exfoliated from 
calcium disilicide~(CaSi$_2$)~\cite{Nakano06a}
and Si nanoribbons have been recently synthesized~\cite{Kara09a, Kara10a, Aufray10a}.
There is a report of the observation of a linear electronic dispersion 
for the Si nanoribbons using angle-resolved photoemission spectroscopy~\cite{DePadova10a}.

Similarly to graphene, there is interest in modifying a silicene sheet
in order to tailor its properties.
For example, studies have shown that Si nanoribbons 
can display semiconducting and magnetic behavior~\cite{Cahangirov09a, Ding09a}.
Hydrogenation of silicene can also open a gap 
(e.g., hydrogenated graphene~(graphane) is an insulator of direct band 
gap~\cite{Sofo07a,Lebegue09b}) and
the resulting compound should be related to polysilane~\cite{Takeda89}.
There are already a few ab initio calculations of related structures.
Indeed, density functional theory (DFT) calculations
within the local density approximation (LDA) found that a Si sheet without
hydrogen is a semiconductor of gap zero~\cite{Lok10a}. Upon hydrogenation,
the Si sheet becomes an semiconductor of indirect band gap~\cite{Lok10a,Takeda89}.
As for hydrogenated Si nanotubes,
a previous density functional tight-binding theory found that hydrogenated
Si nanotubes are semiconductors with a gap that shows little dependence 
on chirality~\cite{Seifert01a}.

It is the purpose of the present work to study 
the band structure of hydrogenated
Si nanosheets and nanotubes 
in more detail than had been done previously
with the ab initio method.
In particular,
($i$) previous work had  focused on structures,
($ii$) the calculations of the hydrogenated Si sheet were all done using LDA,
which suffers from the band-gap problem,
($iii$) a discussion of the wave functions was missing from the earlier work,
and ($iv$) a complete study of the band structures of Si-H nanotubes
has not yet been carried out.  
The most appropriate model is the empirical tight
binding one. 

In this paper, we will first consider a hydrogenated Si sheet
~(also referred as silicane~\cite{Lok10a}).
For the hydrogenated Si sheet, we find that ($i$) it is an semiconductor of indirect band gap of about $2.2$\,eV; ($ii$) 
the lowest point of the conduction bands~(CB) occurs at the M point and the highest point 
of the valence bands~(VB) occurs at the $\Gamma$ point; 
($iii$) the indirect gap is closely related to
the band structure of silicene and it is mainly due to band-filling effect; ($iv$) 
the degeneracy at the Dirac point is 
not lifted by hydrogenating the Si sheet. 
We then consider the electronic structure of hydrogenated single-walled Si nanotubes~(Si-H NTs). 
We find that ($i$) Si-H NTs   
of certain chiralities are semiconductors with co-existing direct and indirect gaps 
of identical magnitudes; ($ii$) the magnitude of the gaps 
is equal to that of the hydrogenated Si sheet. 


\section{Tight-binding model of hydrogenated Si sheets}
\label{sec:TB_model}

We now describe our tight-binding model. 
We consider a $10\times 10 $ Hamiltonian with the 
following Bloch states for Si~($A$ site),
\begin{equation}
\left| A, \alpha, \mathbf{k}\right> = \frac{1}{\sqrt{N}}\sum_{i=1}^Ne^{i\mathbf{k}\cdot(\mathbf{R}_i+\mathbf{\tau}_A )}\left|\mathbf{R}_i,A,\alpha\right>,
\end{equation}
and similarly for the $B$ sites and for H~($C$ and $D$ sites).
Here $\alpha=s,p_x,p_y,p_z$ are L\"{o}wdin orbitals;
$\mathbf{R}_i$ is a lattice vector of the hexagonal lattice;
$\mathbf{\tau}$ is the position vector of the Si and 
H atoms in the unit cell; and $N$ is the number of unit cells. 
For Si atoms, we consider first nearest neighbors located at
\begin{eqnarray*}
&&-{a \over 2 \sqrt{3}}\,\hat\mathbf{x}
                                     +{a \over 2}\,\hat\mathbf{y}\,-\Delta z \,\hat\mathbf{z},
~~~~\frac{a}{\sqrt{3}}\,\hat\mathbf{x}+0\,\hat\mathbf{y}\,-\Delta z \,\hat\mathbf{z},\\
&&-{a \over 2 \sqrt{3}}\,\hat\mathbf{x}
                                     -{a \over 2}\,\hat\mathbf{y}\,-\Delta z\,\hat\mathbf{z},
~~~~~0\,\hat\mathbf{x}+0\,\hat\mathbf{y}+d_{\mbox{\scriptsize{SiH}}}\,\hat\mathbf{z},
\end{eqnarray*}
and second nearest neighbors located at
\begin{eqnarray*}
&&~~~~0\,\hat\mathbf{x}~~
                                     ~+a\,\hat\mathbf{y}\,~~+0\,\hat\mathbf{z},~~~~~~~
0\,\hat\mathbf{x}
                                     ~~~-a\,\hat\mathbf{y}~~+0\,\hat\mathbf{z},\\
&&~~\frac{a \sqrt{3}}{ 2 }\,\hat\mathbf{x}
                                     -\frac{a}{ 2}\,\hat\mathbf{y}+0\,\hat\mathbf{z},~~~~
-\frac{a \sqrt{3}}{2}\,\hat\mathbf{x}
                                     +{a \over 2}\,\hat\mathbf{y}+0\,\hat\mathbf{z},\\
&&~~\frac{a \sqrt{3}}{2}\,\hat\mathbf{x}
                                     +\frac{a}{2}\,\hat\mathbf{y}+0\,\hat\mathbf{z},~~~
-\frac{a \sqrt{3}}{ 2 }\,\hat\mathbf{x}
                                  -\frac{a}{2}\,\hat\mathbf{y}+0\,\hat\mathbf{z}.
\end{eqnarray*}

\begin{figure}[htp]
\begin{centering}
\includegraphics[width=10cm, height=4cm]{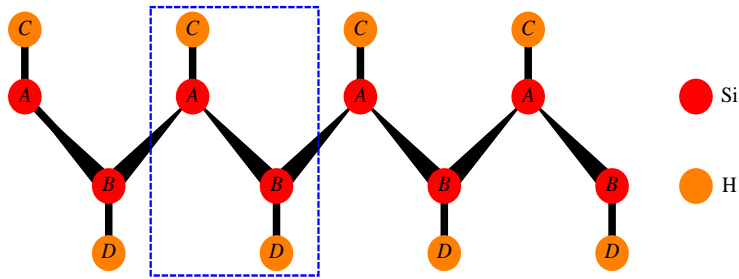}
\caption{Side view of the hydrogenated Si buckled sheet~(silicane). 
Silicon atoms are located at sites $A$ and $B$. 
Hydrogen atoms are located at sites $C$ and $D$. 
The rectangle encloses the unit cell.}
\label{fig:sideview}
\end{centering}
\end{figure}

Here, $\Delta z \simeq 0.19\,a$ is the buckling height; 
$d_{\mbox{\scriptsize{SiH}}}\simeq0.39\,a$ is the Si-H bond length; and $a\simeq 3.820\,${\AA} is the 
lattice constant. These structural parameters are taken
from ab initio calculations~\cite{Lok10a}.
The origin is placed at site A~(figure~\ref{fig:sideview}).
For the hydrogen atoms, we only consider up to first-nearest-neighbor interactions. 
We further assume that there is only a $\sigma$ bond between the Si and H atoms. 
Silicane has a hexagonal lattice with a basis composed of two Si atoms and two hydrogen 
atoms~(figure~\ref{fig:sideview}).

Our tight-binding model is constructed as follows: ($i$)
the Si-Si TB parameters are taken 
from those of bulk Si~($sp^3$ configuration); ($ii$) 
the Si-H TB parameters 
are obtained by using the universal Harrison scaling rule~\cite{Harrison80a}
($iii$) the two-center approximation is used to 
account for the slight distortion of the hydrogenated sheet
from pure $sp^3$ hybridization. 
A simple and fairly good parametrization
of the bulk Si structure is that of Grosso {\it et. al.}
~\cite{Grosso95a}. 
These parameters reproduce fairly well the band structure of bulk Si, in particular 
in the K$\,\Gamma$ direction, which
 is important for the Si nanosheets~\cite{Guzman07a}.
The full list of TB parameters is given in table~\ref{t:TCA-parameters}. 

\Table{\label{t:TCA-parameters}
Tight-binding parameters for the hydrogenated Si sheet. 
Two-center Si-Si parameters are obtained from 
Grosso {\it et al}~\cite{Grosso95a}. 
The H and Si-H parameters are obtained from Harrison~\cite{Harrison80a}
after applying the bond-length scaling for the Si-H parameter. 
All TB parameters are in eV.}
\begin{tabular}{lr} 
\br
Si-Si      &   \\ \mr
$E_{s}(\mbox{Si})$    &   $-4.0497$  \\
$E_{p}$               &   $1.0297$  \\
$(ss\sigma)_{1}^{AB}$ &   $-2.0662$  \\
$(sp\sigma)_{1}^{AB}$ &   $2.1184$   \\
$(pp\sigma)_{1}^{AB}$ &   $3.1866$   \\
$(pp\pi)_{1}^{AB}$    &   $-0.8867$  \\
$(ss\sigma)_{2}^{AA}$ &   $0.0000$   \\
$(sp\sigma)_{2}^{AA}$ &   $0.0000$   \\
$(pp\sigma)_{2}^{AA}$ &   $0.8900$   \\
$(pp\pi)_{2}^{AA}$    &   $-0.3612$  \\ \mr
Si-H    &              \\ \mr
$E_{s}(\mbox{H})$     &   $13.6000$  \\ 
$(sp\sigma)_{1}^{AC}$ &   $6.21820$  \\
\br
\end{tabular}
\endTable


\section{Results and Discussion}
\label{sec:Results_and_discussion}

\begin{figure}[tbp]
\begin{centering}
\includegraphics[width=10cm,height=6cm]{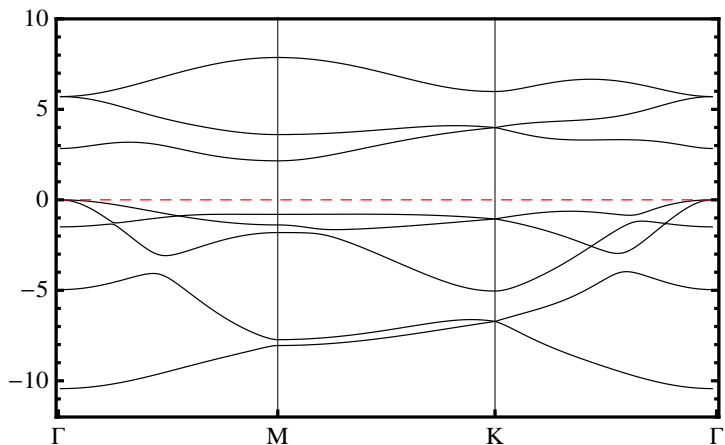}
\caption{Band structure of the hydrogenated Si sheet.
The hydrogenated Si sheet is an semiconductor of indirect 
band gap $E_{g,\mbox{\tiny{ind}}}(\Gamma-M)\simeq 2.2$\,eV. 
The Fermi level is set to zero~(dashed line).}
\label{fig:Si_buckled_sheets}
\end{centering}
\end{figure}

We now present the results obtained from our tight-binding model. 
We first discuss the hydrogenated Si sheet. 
The hydrogenated Si sheet is an semiconductor with 
indirect gap $E_{g,\mbox{\tiny{ind}}}(\Gamma-M)\simeq 2.2\,$eV~(figure~\ref{fig:Si_buckled_sheets}). 
Our result is in between the $2.0\,$eV and $2.5\,$eV reported previously~\cite{Lok10a,Takeda89,Seifert01a}.
The lowest point of the CB is at M and it is 
one-fold degenerate~(figure~\ref{fig:Si_buckled_sheets} 
and table~\ref{t:buckled_sheet_w_h}). The wave function at the M
point of the lowest CB is given as 
follows~(table~\ref{t:buckled_sheet_w_h}): $s$~($43$\%), $p_x$~($31$\%), $p_z$~($20$\%), and $s_H$~($6$\%). 
The highest point in the VB is at $\Gamma$ and it is 
two-fold degenerate~(figure~\ref{fig:Si_buckled_sheets} and table~\ref{t:buckled_sheet_w_h}). 
The wave functions at the $\Gamma$ point of 
the highest VB is pure $p_x$ or pure $p_y$~(table~\ref{t:buckled_sheet_w_h}).
This is the same as was found for graphane~\cite{Sahin10}.

\Table{\label{t:buckled_sheet_w_h}
Wave functions at $\Gamma$, M, and K for the 
hydrogenated Si sheet. The probability densities of the electron states 
at the Si~(H) 
atoms are given by
the addition of the probability densities at sites $A$, $B$~($C$, $D$).}
\begin{tabular}{ccl}
\br
$E_{\mbox{\tiny{Si-H}}}(\Gamma)$\,[eV] &  Degeneracy   & Wave Functions    \\ \mr
$~~\,16.3$   &   $1$        &  $z\,(14\%),s_H\,(86\%)$   \\ 
$~~\,16.0$   &   $1$        &  $z\,(11\%),s_H (89\%)$ \\ 
$~~\,5.7$    &   $2$        &  $x\,(3\%),y\,(97\%);x\,(97\%),y\,(3\%)$ \\ 
$~~\,2.8$    &   $1$        &  $s\,(94\%),z\,(5\%),s_H\,(1\%)$ \\  
$0$           &   $2$        &  $x\,(100\%); y\,(100\%)$   \\
$-1.5$       &   $1$        &  $s\,(4\%),z\,(82\%),s_H\,(14\%)$ \\
$-5.0$       &   $1$        &  $s\,(6\%),z\,(85\%),s_H\,(9\%)$ \\
$-10.4$      &   $1$        &  $s\,(96\%),z\,(4\%)$      \\ \mr

$E_{\mbox{\tiny{Si-H}}}(M)$\,[eV]    &              &           \\ \mr
$~~\,16.6$    &   $1$        &  $z\,(17\%),s_H\,(83\%)$     \\ 
$~~\,16.5$    &   $1$        &  $z\,(15\%),s_H\,(85\%)$ \\  
$~~\,7.9$     &   $1$        &   $y\,(100\%)$\\
$~~\,3.6$     &   $1$        &  $s\,(11\%),x\,(70\%),z\,(14\%),s_H\,(5\%)$   \\  
$~~\,2.2$     &   $1$        &  $s\,(43\%),x\,(31\%),z\,(20\%),s_H\,(6\%)$   \\
$-0.8$        &   $1$        &   $s\,(12\%),x\,(10\%),z\,(66\%),s_H\,(12\%)$     \\
$-1.4$        &   $1$        &  $y\,(100\%)$ \\
$-1.8$        &   $1$        &  $s\,(30\%),x\,(1\%),z\,(60\%),s_H\,(9\%)$ \\ 
$-7.7$        &   $1$        &  $s\,(28\%),x\,(67\%),z\,(5\%)$   \\  
$-8.1$        &   $1$        &  $s\,(77\%),x\,(19\%),z\,(4\%)$    \\ \mr

$E_{\mbox{\tiny{Si-H}}}(K)$\,[eV]  &               &           \\ \mr
$16.6$    &   $2$        &  $z\,(17\%),s_H\,(83\%)$  \\ 
           &              &  $z\,(17\%),s_H\,(83\%)$  \\ 
$6.0$     &   $1$        &  $x\,(50\%),y\,(50\%)$          \\  
$4.0$     &   $2$        &  $s\,(19\%),x\,(31\%),y\,(31\%),z\,(13\%),s_H\,(5\%);$ \\  
           &              &  $s\,(19\%),x\,(31\%),y\,(31\%),z\,(13\%),s_H\,(5\%)$ \\
$-1.1$    &   $2$        &  $s\,(16\%),x\,(3\%),y\,(3\%),z\,(66\%),s_H\,(12\%);$ \\
           &              &  $ s\,(16\%),x\,(3\%),y\,(3\%),z\,(66\%),s_H\,(12\%)$  \\
$-5.0$    &   $1$        &  $x\,(50\%),y\,(50\%)$          \\
$-6.7$    &   $2$        &  $s\,(66\%),x\,(15\%),y\,(15\%),z\,(4\%);$\\ 
           &              &  $s\,(66\%),x\,(15\%),y\,(15\%),z\,(4\%)$  \\ 
\br
\end{tabular}
\endTable

The symmetries of the wave functions at $\Gamma$, M, and K 
for the hydrogenated Si sheet are given in table~\ref{t:buckled_sheet_w_h}. 
Note that the K point~($E_{\mbox{\tiny{Si-H}}}(K)=-1.1\,\mbox{eV},4.0\,\mbox{eV}$) 
is two-fold degenerate~(table~\ref{t:buckled_sheet_w_h}). 
This is expected since fully hydrogenating 
the Si sheet does not break the mirror symmetry of the
honeycomb lattice. 
Another interesting result is that there is little hydrogen $s$ orbital~($s_H$)
in the VB. Thus, the gap ``opening '' is, in reality, 
due to the hydrogen electrons filling up the originally empty
lowest CB of silicene; i.e., the gap is mostly intrinsic to silicene.

Next, we compare the band structures of graphane and silicane. 
The band structures of graphane and silicane differ significantly.  
First, graphane is an insulator of 
direct band gap~\cite{Sofo07a,Lebegue09b}. The gap occurs at $\Gamma$. 
Second, the band gap 
of graphane~($5.4\,$eV) is far larger than that of silicane~\cite{Lebegue09b}.  
This is expected since the 
C-C bond length in graphane is greater than the Si-Si bond length of
silicane~\cite{Lok10a, Sofo07a}. 

\begin{figure}[htp]
\begin{centering}
\includegraphics[width=10cm,height=7cm]{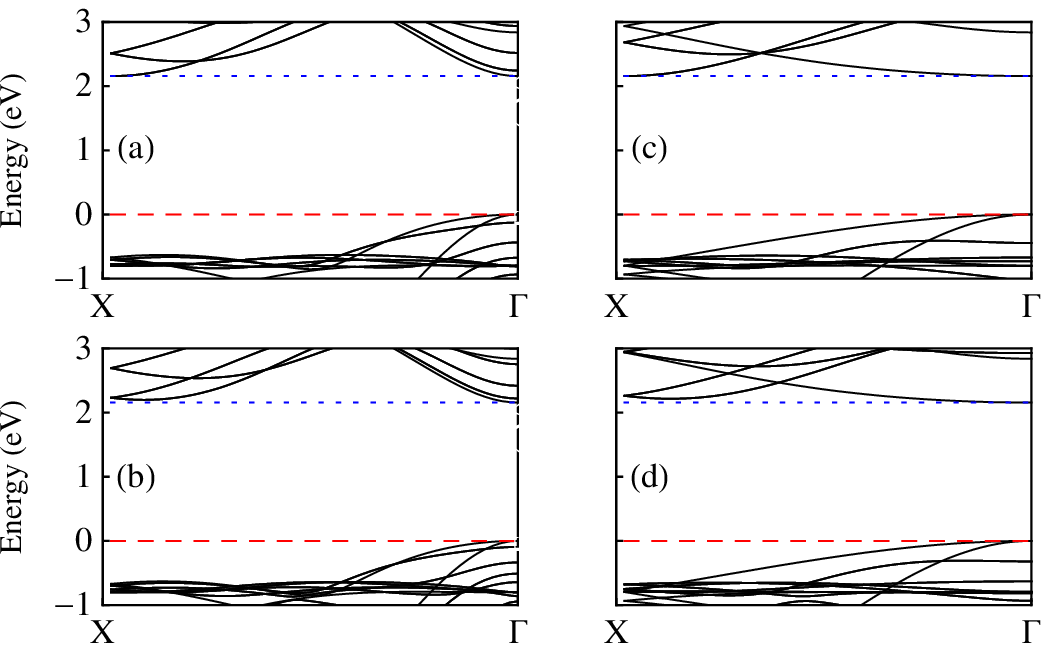}
\caption{Band structure of (a) armchair ($6$,$6$), (b) armchair ($7$,$7$),
 (c) zig-zag ($6$,$0$), (d) zig-zag ($7$,$0$) hydrogenated Si nanotubes. 
Armchair ($n$,$n$) 
nanotubes with $n$ even have simultaneously direct~($\Gamma$) 
and indirect~($\Gamma\,$X) gap equal to that
of the hydrogenated Si sheet. Zig-zag ($n$,$0$) 
nanotubes with $n$ even have simultaneously direct~($\Gamma$) 
and indirect~($\Gamma\,$X) gap
equal to that
of the hydrogenated Si sheet.
For armchair nanotubes 
X$=\pm \pi/a$ and for zig-zag nanotubes X$=\pm \pi/(\sqrt{3}a)$.
The Fermi
level is set to zero~(dashed line). The dotted line in (a) and (c)
is a guide for the eye to show that the lowest conduction
states at $\Gamma$ and X are at the same energy.
The dotted line in (b) and (d)
is a guide for the eye to show that the lowest conduction
states at $\Gamma$ and X are not at the same energy.}  
\label{fig:Si-H NTs}
\end{centering}
\end{figure}

\begin{figure}[htp]
\begin{centering}
\includegraphics[width=10cm,height=12cm]{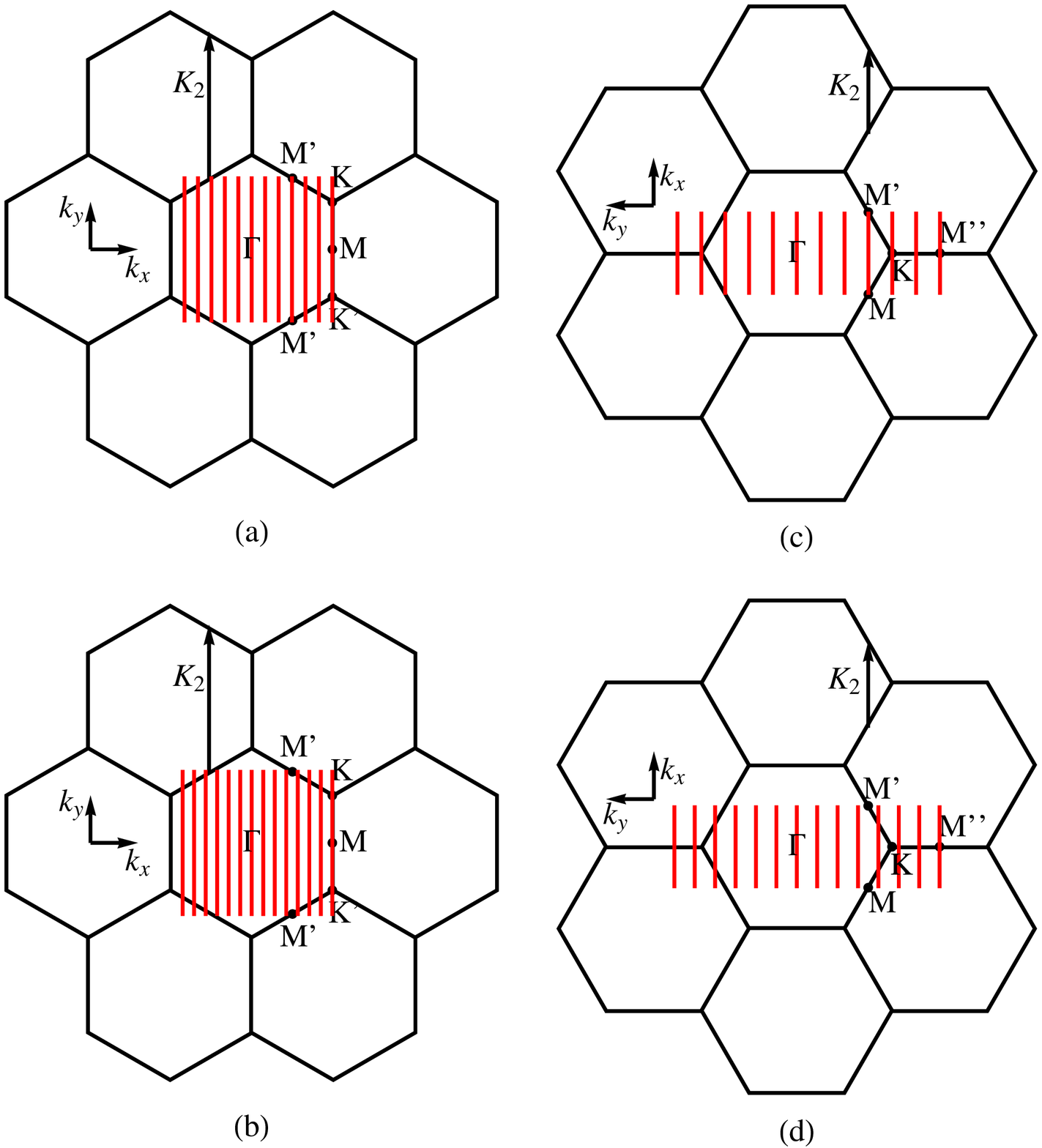}
\caption{Brillouin zone of (a) armchair ($6$,$6$), (b) armchair ($7$,$7$),
 (c) zig-zag ($6$,$0$), (d) zig-zag ($7$,$0$) hydrogenated Si nanotubes. 
Armchair ($n$,$n$) nanotubes with $n$ even have simultaneous direct 
and indirect gaps since its Brillouin zone 
crosses M and M'. Zig-zag ($n$,$0$) 
nanotubes with $n$ even have simultaneous and identical direct and indirect gaps since its Brillouin zone 
crosses M, M' and M''.} 
\label{fig:Si-H-NT-BZ}
\end{centering}
\end{figure}

We now consider the band structure of Si-H NTs. 
Si-H NTs are fully characterized by the chiral 
vector 
$(n,m)$,
where $n,m$ are 
integer numbers~\cite{Saito98a}.  
The band structure of Si-H NTs is obtained from that of 
silicane by quantizing the wave vector $\mathbf{k}$ along the chiral direction~\cite{Saito98a}. 

Armchair ($n$,$n$) and zig-zag ($n$,$0$)
Si-H NTs with $n$ even have concurrent and identical direct and indirect gaps equal to that
of the hydrogenated Si sheet~($2.2\,$eV) (figure~\ref{fig:Si-H NTs}\,(a) and (c)). 
We note that this dual behavior is
different from the so-called mixed character previously reported~\cite{Takeda89}
as the latter was for a nanosheet and was an approximate phenomenon.
Armchair ($n$,$n$) and zig-zag ($n$,$0$)
Si-H NTs with $n$ odd have direct gap equal to that
of the hydrogenated Si sheet~(figure~\ref{fig:Si-H NTs}\,(b) and (d)).  
Therefore, Si-H NTs are semiconductors with either direct gaps, or
with co-existing direct and indirect gaps of identical magnitude.

The electronic behavior of Si-H NTs is explained as follows.
We first consider armchair Si NTs. 
For armchair Si-H NTs~$(n,n)$, quantization of the 
wave vector $\mathbf{k}$ along the chiral direction gives their 
Brillouin zone~(BZ)~\cite{Saito98a},
\begin{equation}
\label{eq:armchair_BZ}
\mathbf{k}=\frac{2\pi}{a}\frac{q}{\sqrt{3}n}\,\hat{ \mathbf{k}}_x+k_{y}\,\hat{ \mathbf{k}}_y,
\end{equation}
where, $q=-n+1,...,0,...,n$ and $-\pi/a < k_y < \pi/a$. 
Armchair ($n$,$n$) Si-H-NTs 
with $n$ even have simultaneous and identical direct and indirect gaps 
since their BZ, equation~(\ref{eq:armchair_BZ}), always crosses M and 
M'~(figure~\ref{fig:Si-H-NT-BZ}~(a)),
\begin{eqnarray*}
{\bf M}= \frac{2\pi}{a}\frac{1}{\sqrt{3}}\, \hat{\mathbf{k}}_x,~
{\bf M'}= \frac{2\pi}{a}\left(\frac{1}{2\sqrt{3}}\,\hat{\mathbf{k}}_x\pm \frac{1}{2}\,\hat{\mathbf{k}}_y\right).
\end{eqnarray*}
Armchair ($n$,$n$) Si-H NTs with $n$ odd have only a direct gap
since their BZ always crosses M but not M'~(figure~\ref{fig:Si-H-NT-BZ}~(b)). 

We now consider zig-zag Si-H NTs.
For zig-zag Si-H NTs~$(n,0)$, quantization of the wave vector $\mathbf{k}$
along the chiral direction gives their BZ~\cite{Saito98a},
\begin{equation}
\label{eq:zig-zag_BZ}
\mathbf{k}=k_{x}\,\hat{ \mathbf{k}}_x+\frac{2\pi}{a}\frac{q}{n}\,\hat{ \mathbf{k}}_y,
\end{equation}
where, $q=-n,...,0,...,n-1$ and $-\pi/(\sqrt{3}a) < k_x < \pi/(\sqrt{3}a)$.
Zig-zag ($n$,$0$) Si-H NTs 
with $n$ even have simultaneous and identical direct and indirect gaps
since their BZ, equation~(\ref{eq:zig-zag_BZ}), 
always crosses M, M' and M''~(figure~\ref{fig:Si-H-NT-BZ}~(c)),
\begin{eqnarray*}
\hspace{-1cm}
{\bf M}=\frac{2\pi}{a}\left(-\frac{1}{2\sqrt{3}}\,\hat{\mathbf{k}}_x-\frac{1}{2}\,\hat{\mathbf{k}}_y\right),~
{\bf M'}=\frac{2\pi}{a}\left(\frac{1}{2\sqrt{3}}\,\hat{\mathbf{k}}_x-\frac{1}{2}\,\hat{\mathbf{k}}_y\right),~
{\bf M''}=-\frac{2\pi}{a}\,\hat{\mathbf{k}}_y.
\end{eqnarray*}

Zig-zag ($n$,$0$) Si-H NTs with $n$ odd have only direct gap
since their BZ always crosses M'' but not M, M'~(figure~\ref{fig:Si-H-NT-BZ}~(d)). 
The above results on the gap dependence on chirality are much more
complex than the Hamada rule~\cite{Hamada92a} 
obeyed by non-hydrogenated carbon and silicon nanotubes~\cite{Guzman07a}.

We now compare our Si-H NTs results to those of Ref.~\cite{Seifert01a}.
Ref.~\cite{Seifert01a} presents 
density functional tight-binding~(DFTB) 
calculations of the Si-H NT gap. 
The DFTB calculation finds that Si-H NTs are semiconductors whose 
gap has little dependence on chirality. The gap of 
Si-H NTs rapidly converges to the gap of their hydrogenated Si sheet
($\simeq 2.5$\,eV).
No reference
is made as to whether the gap is direct or indirect.
Even though the electronic behavior 
of our TB model is in fairly good agreement with that
of DFTB, one important difference is that
our TB calculation did not find a dependence of 
the size of the gap at all.
We attribute this difference
to curvature effects, which are neglected in our model.

\section{Conclusions}
\label{sec:Conclusions}

In summary, we performed a tight-binding calculation of the band structure 
of fully hydrogenated Si nanostructures.  For the hydrogenated 
Si sheet~(silicane), we find that it is a semiconductor with 
indirect band gap of about $2.2\,$eV. The indirect band gap occurs 
between the M and $\Gamma$ points. 
Our study of the wave functions reveals that the gap is
closely related to the band structure of silicene
and is primarily due to a band-filling effect rather than to
a gap opening.
For Si-H NTs, we find that they are semiconductors with either a direct gap 
or with co-existing direct and indirect band gaps depending on chirality. 
Therefore, Si-H NTs do not follow Hamada's rule.
This shows that hydrogenated Si nanotubes can have properties distinct from
carbon nanotubes.


\end{document}